\documentclass[equations,11pt]{article}\relax
\usepackage{amssymb,amsmath,amsthm} \relax
\usepackage{amsmath}
\usepackage{eucal}
\renewcommand{\theequation}{\arabic{equation}}
\begin{document}
\bibliographystyle{plain}
\def\m@th{\mathsurround=0pt}
\mathchardef\bracell="0365 
\def\upbrall{$\m@th\bracell$}
\def\undertilde#1{\mathop{\vtop{\ialign{##\crcr
    $\hfil\displaystyle{#1}\hfil$\crcr
     \noalign
     {\kern1.5pt\nointerlineskip}
     \upbrall\crcr\noalign{\kern1pt
   }}}}\limits}
\def\theequation{\arabic{section}.\arabic{equation}}

\newcommand{\pii}{P$_{{\rm\small II}}$}  
\newcommand{\pvi}{P$_{{\rm\small VI}}$}   
\newcommand{\ptf}{P$_{{\rm\small XXXIV}}$}   
\newcommand{\diffE}{O$\triangle$E}   

\newcommand{\pp}{\partial}
\newcommand{\ar}{\alpha}
\newcommand{\aar}{\bar{a}}
\newcommand{\bb}{\beta}
\newcommand{\gm}{\gamma}
\newcommand{\Gm}{\Gamma}
\newcommand{\en}{\epsilon}
\newcommand{\ven}{\varepsilon}
\newcommand{\dd}{\delta}
\newcommand{\sg}{\sigma}
\newcommand{\kp}{\kappa}
\newcommand{\ld}{\lambda}
\newcommand{\bmu}{\bar{\mu}}
\newcommand{\vf}{\varphi}
\newcommand{\Ups}{\Upsilon}
\newcommand{\oa}{\omega}
\newcommand{\hf}{\frac{1}{2}}
\newcommand{\bea}{\begin{eqnarray}}
\newcommand{\eea}{\end{eqnarray}}
\newcommand{\bse}{\begin{subequations}}
\newcommand{\ese}{\end{subequations}}
\newcommand{\nn}{\nonumber}
\newcommand{\bR}{\bar{R}}
\newcommand{\bP}{\bar{\Phi}}
\newcommand{\bS}{\bar{S}}
\newcommand{\bu}{{\boldsymbol u}}
\newcommand{\bt}{{\boldsymbol t}}
\newcommand{\bm}{{\boldsymbol m}}
\newcommand{\boa}{{\boldsymbol \omega}}
\newcommand{\bet}{{\boldsymbol \eta}}
\newcommand{\bW}{\bar{W}}
\newcommand{\sn}{{\rm sn}}
\newcommand{\wh}{\widehat}
\newcommand{\ol}{\overline}
\newcommand{\wt}{\widetilde}
\newcommand{\ut}{\undertilde}
 \newcommand{\bU}{\bf U}
 \newcommand{\pl}{\partial}
 \newcommand{\ddp}{\frac{\partial}{\partial p}}
 \newcommand{\ddq}{\frac{\partial}{\partial q}}
 \newcommand{\ddr}{\frac{\partial}{\partial r}}
 \newcommand{\Ld}{{\bf \Lambda}}
 \newcommand{\tLd}{\,^{t\!}{\bf \Lambda}}
 \newcommand{\I}{{\bf I}}
 \newcommand{\tII}{\,^{t\!}{\bf I}}
 \newcommand{\tuk}{\,^{t\!}{\bf u}_k}
 \newcommand{\tul}{\,^{t\!}{\bf u}_\ell}
 \newcommand{\tcl}{\,^{t\!}{\bf c}_{\ell}}
 \newcommand{\ssk}{\sigma_{k^\prime}}
 \newcommand{\ssl}{\sigma_{\ell^\prime}}
 \newcommand{\ddint}{\int_\Gamma d\ld(\ell) }
 \def\hypotilde#1#2{\vrule depth #1 pt width 0pt{\smash{{\mathop{#2}
 \limits_{\displaystyle\widetilde{}}}}}}
 \def\hypohat#1#2{\vrule depth #1 pt width 0pt{\smash{{\mathop{#2}
 \limits_{\displaystyle\widehat{}}}}}}
 \def\hypo#1#2{\vrule depth #1 pt width 0pt{\smash{{\mathop{#2}
 \limits_{\displaystyle{}}}}}}

\newtheorem{theorem}{Theorem}[section]
\newtheorem{lemma}{Lemma}[section]
\newtheorem{cor}{Corollary}[section]
\newtheorem{prop}{Proposition}[section]
\newtheorem{definition}{Definition}[section]
\newtheorem{conj}{Conjecture}[section]
 
\begin{center}
{\large{\bf On a Schwarzian PDE associated with the KdV Hierarchy}} 
\vspace{1.2cm}

Frank Nijhoff\vspace{.15cm} \\
{\it Department of Applied Mathematics\\
The University of Leeds, Leeds LS2 9JT, UK}\\
\vspace{.5cm}
Andrew Hone and Nalini Joshi\vspace{.15cm}\\
{\it Department of Pure Mathematics\\
The University of Adelaide\\
Adelaide, Australia 5005}
\vspace{.5cm}

\today
\end{center}
\vspace{1.4cm}
 
\centerline{\bf Abstract}  
\vspace{.2cm}

We present a novel integrable non-autonomous partial differential equation 
of the Schwarzian type, i.e. invariant under M\"obius transformations, 
that is related to the Korteweg-de Vries hierarchy. In fact, this 
PDE can be considered as the generating equation for the entire 
hierarchy of Schwarzian KdV equations. We present its Lax pair, establish 
its connection with the SKdV hierarchy, its Miura relations to 
similar generating PDE's for the modified and regular KdV hierarchies 
and its Lagrangian structure. 
Finally we demonstrate that its similarity reductions lead to the 
{\it full} Painlev\'e VI equation, i.e. with four arbitary parameters. 
\vfill
 
 
\setcounter{page}{0}
\pagebreak
 
\section{Introduction} 
\setcounter{equation}{0} 
In this note we introduce a non-autonomous 1+1-dimensional nonlinear evolution 
equation for the variable $z(s,t)$, namely: 
\bea\label{eq:SPDE}  
z_{sstt}&=& z_{sst}\Bigl( \frac{z_{st}}{z_s}+\frac{z_{tt}}{z_t}\Bigr) 
+ z_{stt}\Bigl( \frac{z_{st}}{z_t}+\frac{z_{ss}}{z_s}\Bigr) 
- z_{st}\Bigl( \frac{z_{st}z_{ss}}{z_s^2} + 
\frac{z_{st}z_{tt}}{z_t^2} + \frac{z_{ss}z_{tt}}{z_sz_t} \Bigr) \nn \\ 
&&+\frac{1}{s-t}\Bigl[ \frac{s}{t}\Bigl( z_{sst} -
\frac{z_{st}z_{ss}}{z_s} -\frac{1}{2}\frac{z_{st}^2}{z_t}\Bigr) -  
\frac{t}{s}\Bigl( z_{stt} - \frac{z_{st}z_{tt}}{z_t} -
\frac{1}{2}\frac{z_{st}^2}{z_s}\Bigr) \Bigr] \nn \\  
&&-\frac{1}{(s-t)^2}\Bigl[ n^2\frac{s^2}{t^2}\frac{z_s}{z_t}
\Bigl( z_{st} - \frac{z_{s}z_{tt}}{z_t}\Bigr) +  
m^2\frac{t^2}{s^2}\frac{z_t}{z_s}\Bigl( z_{st} -
\frac{z_{t}z_{ss}}{z_s}\Bigr) \Bigr] \nn \\ 
&&-\frac{1}{2}\frac{1}{(s-t)^3}\Bigl[ n^2\frac{s}{t}z_s
\Bigl( 1+\frac{(4t-3s)s}{t^2} \frac{z_s}{z_t}\Bigr) -  
m^2\frac{t}{s}z_t\Bigl( 1+\frac{(4s-3t)t}{s^2} \frac{z_t}{z_s}\Bigr) 
\Bigr] \   . 
\eea 
We will refer to this equation simply as {\it the} Schwarzian PDE (SPDE). 
It is straightforward to verify that it is invariant under M\"obius 
transformations
\[ z(s,t)\   \ \mapsto\   \ \frac{\ar z(s,t)+\bb}{\gm z(s,t)+\dd}  \   , \] 
with constant $\ar$,$\bb$,$\gm$,$\dd$. 
It is straightforward to verify also that the equation (\ref{eq:SPDE}) 
arises as the Euler-Lagrange equations from the following 
Lagrangian density:
\begin{equation} \label{eq:Lagr} 
\mathcal{L}[z(s,t)] = 
\frac{1}{2}\frac{(s-t)}{st}\frac{z_{st}^2}{z_s z_t} 
+ \frac{1}{2(s-t)}\Bigl( n^2\frac{s}{t^3}\frac{z_s}{z_t} + 
m^2\frac{t}{s^3}\frac{z_t}{z_s} \Bigr)\  .   
\end{equation} 
Note that the lagrange density $\mathcal{L}$ depends explicitly 
on the independent variables $s$ and $t$. 
We further assert the following properties of the SPDE:
\begin{enumerate}
\item The SPDE (\ref{eq:SPDE}) is integrable in the conventional 
soliton sense: there exists a Lax pair and an infinity of conservation 
laws. 
\item The SPDE (\ref{eq:SPDE}) is part of a {\it compatible} 
parameter-family of PDEs, meaning that the same dependent variable $z$ 
obeys eq. (\ref{eq:SPDE}) simultaneously for all possible pairs   
of parameters $n$,$m$ (with each parameter being associated with 
a different independent variable). We will make this statement more 
precise in what follows. 

\item This infinite family of PDEs is equivalent to the 
Schwarzian KdV hierarchy via expansions on the {\it independent} variables. 
\item As indicated already the system carries a Lagrangian structure. We 
expect that from this Lagrangian one can infer that 
the system is multi-Hamiltonian, although we will not discuss this 
point further here. 
\item There exists a Miura Chain, i.e. a short sequence of differential 
relations connecting the SPDE with other generating PDEs sharing the above 
properties with the SPDE. 
\item The similarity reduction of the SPDE under scaling symmetry reduces 
the SPDE to the full sixth Painlev\'e equation (PVI) with four 
arbitrary parameters. 
\item The SPDE has a fully discrete counterpart which is integrable in its 
own right. 
\end{enumerate}

\section{Properties of the SPDE and related generating PDEs}
\setcounter{equation}{0} 
We now demonstrate or give arguments for the properties that we 
listed in the Introduction. 

\subsection*{Lax pair} 

The following linear overdetermined set of 
equations constitute a Lax pair for the SPDE:  
\bse\label{eq:Lax}\bea
2t\vf_t &=& n\left(\begin{array}{cc} 1&0\\ 0&a\end{array}\right) \vf + 
\frac{1}{\kp-t}\left(\begin{array}{cc} -\kp n(1+a)&2t^2z_t\\ 
\kp n^2(1-a^2)/(2tz_t)&-nt(1-a)\end{array}\right) \vf, \label{eq:Laxa}\\ 
2s\vf_s &=& m\left(\begin{array}{cc} 1&0\\ 0&b\end{array}\right) \vf + 
\frac{1}{\kp-s}\left(\begin{array}{cc} -\kp m(1+b)&2s^2z_s\\ 
\kp m^2(1-b^2)/(2sz_s)&-ms(1-b)\end{array}\right)\vf.  \label{eq:Laxb} 
\eea\ese 
In the above $a=a(s,t)$ and $b=b(s,t)$ are auxiliary variables. The 
compatibility conditions arising from the consistency of cross-differentiating 
(\ref{eq:Laxa}) and (\ref{eq:Laxb}) are 
\bse\label{eq:zab}\bea
&& nsa_s=mtb_t=\frac{1}{2(s-t)}\left[ n^2\frac{s^2}{t}\frac{z_s}{z_t}(1-a^2)
- m^2\frac{t^2}{s}\frac{z_t}{z_s}(1-b^2) \right]  \   , \label{eq:zaba}\\ 
&& stz_{st}=\frac{mt^2z_t b-ns^2z_s a}{s-t}\    . \label{eq:zabb} 
\eea\ese 
The SPDE (\ref{eq:SPDE}) follows immediately from the relations 
(\ref{eq:zaba}) and (\ref{eq:zabb}). In fact, differentiating 
(\ref{eq:zabb}) with respect to $s$ and with respect to $t$ 
yields relations for $a_t$ and $b_s$, whilst (\ref{eq:zaba}) 
gives $a_s$ and simultaneously $b_t$. Thus, with these expressions 
we can cross-differentiate, e.g. $\pl_t a_s=\pl_s a_t$, to obtain 
an equation where we can use the previous obtained expressions to 
to eliminate all $a$'s and $b$'s (miraculously, it turns out 
that the single remaining $a$ or $b$ drops out entirely at the end). 

We note that the Lax pair (\ref{eq:Lax}) falls in the general class 
of Lax pairs of {\it non-isospectral type}\footnote{This does not 
mean that the spectral parameter $\kp$ depends on $s$,$t$: $\kp$ is 
constant w.r.t. the independent variables. However, the poles in the 
Lax matrices depend explicitly on $s$ and $t$. } that were treated in 
\cite{Burtsev}. However, although general families of Lax equations 
were postulated in that paper, the precise reduction which leads to 
(\ref{eq:Lax}) was not investigated there. 

\subsection*{Consistency and Integrability} 

The SPDE (\ref{eq:SPDE}) is compatible with itself in the following 
sense: if one considers for one and the same function $z=
z(t_1,t_2,t_3,\dots)$ of many independent variables 
$t_1,t_2,t_3,\dots$ copies of the equation (\ref{eq:SPDE}) in terms of any two 
distinct variables $t_i,t_j$ replacing $s$ and $t$ in(\ref{eq:SPDE}), 
each associated with its own 
parameters $n_i,n_j$ replacing $n$ and $m$ in (\ref{eq:SPDE}), then 
all these copies of the SPDE on one and the same function $z$ are 
consistent. Thus, 
we have effectively an infinite family of commuting flows all given 
by the same equation, but with different parameters: in other 
words the SPDE represents a {\it parameter-family of consistent PDE's}. 

This statement can be verified by direct calculation, but 
since (\ref{eq:SPDE}) is not in evolution form the verification 
is most easily done using the Lax pair. In fact, suppose we augment 
the system (\ref{eq:Lax}) with a third part, i.e. an equation 
similar to (\ref{eq:Laxa}) and (\ref{eq:Laxb}) but now in terms 
of a third independent variable ($u$ say) associated with a new 
parameter $\ell$ (instead of $n$ or $m$) and a third auxiliary variable 
$c$. Obviously the compatibility 
relations with the original members of the Lax pair, namely with 
(\ref{eq:Laxa}) or (\ref{eq:Laxb}) lead to extra relations of 
the same type as (\ref{eq:zab}) involving now relations for $a_u$, 
$b_u$, $c_s$, $c_t$ as wel as $z_{su}$ and $z_{tu}$. We have to 
check the consistency of these relations, i.e. the validity of 
expressions such as  ~$\pl_u a_s=\pl_s a_u$~  and  
~$\pl_u z_{st}=\pl_t z_{su}=\pl_s z_{tu}$~. This follows by tedious but 
straightforward  direct calculation. 

As is clear from the variational equations from the Lagrangian 
(\ref{eq:Lagr}), namely 
\begin{equation}\label{eq: varia}
\frac{\dd\mathcal{L}}{\dd z}= \frac{\pl^2}{\pl s\pl t}\left( 
\frac{\pl\mathcal{L}}{\pl z_{st}} \right) - 
\frac{\pl}{\pl s}\left( \frac{\pl\mathcal{L}}{\pl z_{s}} \right) -
\frac{\pl}{\pl t}\left( \frac{\pl\mathcal{L}}{\pl z_{t}} \right) 
=0\   , 
\end{equation} 
the SPDE can be written as a conservation law. Since we know from 
the above argument that the SPDE can be imposed on $z=z(s,t,u,\dots)$ 
in as many independent variables as we want (each associated with its 
own parameter $n$,$m$,$\ell$,\dots), we derive in this way an 
infinity of conservation laws all for the same object $z$. What is 
potentially nontrivial is to rewrite these conservation laws 
in terms of only one pair of preferred variables $s$ and $t$ say. 
It is only at this point that possible nonlocalities in the 
explicit expressions for the conservation laws may occur. We will 
not discuss this point here in detail any further, but postpone 
this discussion to future investigations of the SPDE. 

\subsection*{Connection with SKdV Hierarchy} 

Using the expansion of the independent variables: 
\begin{equation}\label{eq:expand}
\frac{\pl}{\pl t}=-\frac{n}{t^{1/2}}\sum_{j=1}^\infty \frac{1}{t^j} 
\frac{\pl}{\pl x_j}\    \ ,\    \ 
\frac{\pl}{\pl s}=-\frac{m}{s^{1/2}}\sum_{j=1}^\infty \frac{1}{s^j} 
\frac{\pl}{\pl x_j}\   , 
\end{equation}
where $x_1,x_2,\dots$ is the infinite sequence of higher times 
associated with the KdV hierarchy, we can expand the Lagrangian 
(\ref{eq:Lagr}) in powers of $s$ and $t$ as follows: 
\bea 
\mathcal{L}&=& \frac{1}{2} \frac{mn}{(st)^{3/2}}\left\{ 
\frac{s+t}{s-t} + \left( \frac{1}{s}-\frac{1}{t}\right) 
\left[ \frac{z_{x_2}}{z_{x_1}}-\frac{z_{x_1x_1}^2}{z_{x_1}^2} \right] 
\right. \nn\\ 
&&~~ + \left. \left( \frac{1}{s^2}-\frac{1}{t^2}\right) 
\left[ \frac{z_{x_3}}{z_{x_1}}-2\frac{z_{x_1x_1}z_{x_1x_2}}{z_{x_1}^2} 
+ \frac{z_{x_1x_1}^2 z_{x_2}}{z_{x_1}^3}\right] + \dots \right\}\  .   
\eea 
Thus, order by order we obtain a sequence of Lagrangians which turns 
out to be exactly the Lagrangians for the SKdV hierarchy,   
the first 
equations in the sequence reading  
\bse \label{eq:SKdVhier} 
\begin{equation}\label{eq:SKdV} 
\frac{z_{x_2}}{z_{x_1}}=\{ z,x_1 \} \equiv \frac{z_{x_1x_1x_1}}{z_{x_1}}-
\frac{3}{2} \frac{z_{x_1x_1}^2}{z_{x_1}^2}\  , 
\end{equation}
which is the Schwarzian KdV equation, 
cf. \cite{weiss:II,Mokh}, and 
\begin{equation}\label{eq:HiSKdV} 
\frac{z_{x_3}}{z_{x_1}}= 2 \frac{\pl^2}{\pl x_1^2}\left( 
\frac{z_{x_2}}{z_{x_1}}\right) +\frac{3}{2}
\frac{z_{x_2}^2}{z_{x_1}^2}\  , 
\end{equation} \ese 
which is the first higher-order SKdV equation.  
(Strictly speaking the sequence of Lagrangians appearing in 
(\ref{eq:SKdVhier}) yield the $x_{1}$ derivatives 
of this hierarchy of equations.) The expansions
(\ref{eq:expand}) amount exactly to the transition from the higher 
time variables in the hierarchy to so-called Miwa coordinates, 
\cite{Miwa}. 
We mention that the notion of a generating PDE for a hierarchy 
of integrable nonlinear evolution equations, and their corresponding 
Lagrangians, is in spirit the same as the idea of ``compounding 
hierarchies'' that was put forward some years ago by one of the 
authors, cf. \cite{Oberwol}. We refer also to \cite{KMA} for 
related issues concerning the Baker-Akhiezer function for the 
KP hierarchy in terms of Miwa coordinates. 

\subsection*{Miura Chain} 

The different PDEs in the Miura chain are related via the sequence: 
\vspace{.1cm}
\begin{center}
{\sf Schwarzian PDE }\hspace{.3cm} $\longrightarrow$ \hspace{.3cm}
{\sf Modified PDE }\hspace{.3cm} $\longrightarrow$ \hspace{.3cm}
{\sf Regular PDE }\hspace{.3cm}
\end{center}

The  
MPDE (Modified PDE)   
is the equation for a variable $v(s,t)$ that is connected to 
the auxiliary variables $a(s,t)$ and $b(s,t)$ via:
\begin{equation}\label{eq:vab}
na=-2t\pl_t \log v\      \ ,\     \ 
mb=-2s\pl_s\log v\   , 
\end{equation}
and the relevant equation can be obtained from (\ref{eq:zaba}) in the 
form
\bse\label{eq:MPDE} 
\begin{equation}\label{eq:vst}
\pl_s\pl_t\log v= \frac{nm}{4st(s-t)}\left[ t(1-a)(1+b)Y-
s(1+a)(1-b)\frac{1}{Y}\right]\   ,  
\end{equation}
together with 
\bea\label{eq:Yst}
2st\pl_s\pl_t\log Y &=& ns\pl_s\left[(1-Y)\frac{2t Y +(s-tY)(1+a)}{(t-s)Y}
\right] \nn\\ 
&& ~ - mt\pl_t\left[(1-Y)\frac{2s-(s-tY)(1+b)}{(t-s)Y}\right]  . 
\eea
\ese 
Solving from the quadratic equation (\ref{eq:vst}) for 
\begin{equation}\label{eq:Y}
Y\equiv \frac{m}{n}\frac{t z_t}{s z_s}\frac{(1-b)}{(1-a)} 
\end{equation}
in terms of $v$ and its derivatives (which enter in $a$ and $b$ 
via (\ref{eq:vab})) and substituting the result into eq. 
(\ref{eq:Yst}) we obtain a complicated-looking PDE in terms 
of $v$, of second degree in the highest derivative $v_{sstt}$: that equation 
is the MPDE, 
but because of its length we omit the explicit formula. 

Finally, the Regular PDE, by which we mean the PDE that 
upon expansion yields actually the entire hierarchy of KdV equations, 
is given by  
\bea\label{eq:UPDE}  
U_{sstt}&=& U_{sst}\Bigl( \frac{1}{s-t}+\frac{U_{st}}{U_s}+
\frac{U_{tt}}{U_t}\Bigr) 
+ U_{stt}\Bigl( \frac{1}{t-s}+\frac{U_{st}}{U_t}+
\frac{U_{ss}}{U_s}\Bigr) 
- U_{st} \frac{U_{ss}U_{tt}}{U_sU_t}  \nn \\ 
&& + U_{ss}\Bigl( \frac{m^2}{(s-t)^2}\frac{U_t^2}{U_s^2}
-\frac{U_{st}^2}{U_s^2} - \frac{1}{s-t} 
\frac{U_{st}}{U_s}\Bigr) 
+ U_{tt}\Bigl( \frac{n^2}{(s-t)^2}\frac{U_s^2}{U_t^2}
-\frac{U_{st}^2}{U_t^2} + \frac{1}{s-t} 
\frac{U_{st}}{U_t}\Bigr) \nn \\ 
&& + \frac{m^2}{2(s-t)^3}\frac{U_t}{U_s}\Bigl( U_s+U_t+2(t-s)U_{st}
\Bigr)   
- \frac{n^2}{2(s-t)^3}\frac{U_s}{U_t}\Bigl( U_s+U_t+2(s-t)U_{st}
\Bigr)  \nn \\ 
&& + \frac{1}{2(s-t)} U_{st}^2\Bigl( \frac{1}{U_s}-
\frac{1}{U_t}\Bigr) \   . 
\eea 
This equation derives from the Lagrangian: 
\begin{equation}\label{eq:ULagr} 
\mathcal{L}[U(s,t)]= \frac{1}{2}(s-t)\frac{U_{st}^2}{U_sU_t} 
+\frac{1}{2(s-t)}\Bigl( n^2\frac{U_s}{U_t} + 
m^2\frac{U_t}{U_s} \Bigr)\   .  
\end{equation} 
It is remarkable that the only apparent difference between the 
Lagrangians (\ref{eq:Lagr}) and (\ref{eq:ULagr}) resides in the 
way the variables $s$ and $t$ enter. 

The PDE (\ref{eq:UPDE}) arises from the following Lax pair: 
\bse\label{eq:ULax}\bea
\phi_t &=& \frac{n}{2t^{1/2}}\left(\begin{array}{cc} 0&0\\ 1&0
\end{array}\right) \phi - 
\frac{1}{\kp-t}\left(\begin{array}{cc} n-A U_t&U_t\\ 
A(n-A U_t)& A U_t\end{array}\right) \phi \label{eq:ULaxa}\\ 
\phi_s &=& \frac{m}{2s^{1/2}}\left(\begin{array}{cc} 0&0\\ 1&0
\end{array}\right) \phi - 
\frac{1}{\kp-s}\left(\begin{array}{cc} m-B U_s&U_s\\ 
B(m-B U_s)& B U_s\end{array}\right) \phi \label{eq:ULaxb} 
\eea\ese 
where $A(s,t)$ and $B(s,t)$ are some auxiliary variables. The 
non-isospectral Lax pair (\ref{eq:ULax}) 
is gauge-equivalent to the Lax representation (\ref{eq:Lax}) 
via the gauge transformation 
\begin{equation}\label{eq:gauge}
\vf=\left( \begin{array}{cc} S &v\\ \kp/v & 0
\end{array} \right) \phi\   ,  
\end{equation} 
in which $S$ is yet another auxiliary variable which is determined 
by the consistency relation of (\ref{eq:gauge}) with both Lax 
relations (\ref{eq:Lax}) and (\ref{eq:ULax}). 

The compatibility conditions between (\ref{eq:ULaxa}) and 
(\ref{eq:ULaxb}) yield the coupled relations 
\bse\label{eq:ABU}\bea 
U_{st} &=& \frac{1}{s-t} \Bigl[ mU_t-nU_s+2(A-B)U_sU_t\Bigr] 
\label{eq:ABUa}\\ 
A_s&=&\frac{m}{2s^{1/2}}+\frac{1}{t-s}\Bigl[ (A-B)^2U_s+m(A-B)
\Bigr] \label{eq:ABUb}\\ 
B_t&=&\frac{n}{2t^{1/2}}+\frac{1}{s-t}\Bigl[ (A-B)^2U_t-n(A-B)
\Bigr]\  .  \label{eq:ABUc} 
\eea \ese 
Combining (\ref{eq:ABUb}) and (\ref{eq:ABUc}) we obtain a coupled 
equation for $A-B$ and $U$, which yields the PDE (\ref{eq:UPDE}) 
after eliminating $A-B$ by solving for the latter in (\ref{eq:ABUa}).  

The Miura transformation between the Regular PDE (\ref{eq:UPDE}) 
and the Modified PDE given by the system (\ref{eq:MPDE})  can be 
constructed on the basis of the various intermediate relations between 
the objects $U$,$Y$,$a$,$b$,$A$ and $B$ and is given by 
\begin{equation}\label{eq:Miura} 
\frac{(s-t)U_{st}-mU_t-nU_s}{(s-t)U_{st}+mU_t+nU_s}= 
\frac{(n+2tw_t)U_s+(m-2sw_s)U_t}{(n-2tw_t)U_s+(m+2sw_s)U_t}\   , 
\end{equation} 
where $w\equiv\log v$. 

Finally, we remark that the analogue of the famous Cole-Hopf 
transformation that interpolates between the SKdV equation and the 
MKdV equation is given by an implicit transformation that interpolates 
between the SPDE (\ref{eq:SPDE}) and the Modified PDE (\ref{eq:MPDE}) 
and that can be straightforwardly inferred from (\ref{eq:Y}) in 
combination with (\ref{eq:vst}).

\subsection*{Connection with PVI}

The similarity reduction of the SPDE is obtained via the constraint
\begin{equation}
\label{eq:spdeconstr}
\mu z+tz_t+sz_s = 0 \   .
\end{equation}
Using the constraint to elminate the derivatives with 
respect to one of the 
independent variables ($s$, say) and introducing the new variable 
\footnote{Alternatively, the similarity reduction can be obtained by 
explicitly solving the constraint (\ref{eq:spdeconstr}) as 
~$z(s,t)=(st)^{-\mu/2} Z(s/t)$~. This leads to a third order 
equation for $Z(s/t)$, namely the Schwarzian PVI 
which we found in \cite{NJH} via a  
different approach.} 
\[ W(t)= 1+ \frac{t}{\mu}\frac{z_t}{z}  \] 
we obtain from the SPDE a third-order ODE for $W$ which 
reads: 
\bea
&& W'''= 2\left( \frac{1}{W} + \frac{1}{W-1}\right) W'W'' 
- 2\left( \frac{1}{t} + \frac{1}{t-s}\right) W''  \nn \\ 
&& ~~~~ - \left( \frac{1}{W^2} + \frac{1}{(W-1)^2} + 
\frac{1}{W(W-1)}\right) W^{\prime 3} 
-\frac{(11t-5s)W+2s-5t}{2(s-t)tW(W-1)} W^{\prime 2}  \nn \\ 
&& ~~~~ +\frac{n^2s^2W^3-m^2t^2(W-1)^3 + \mu^2(s-t)^2W^3(W-1)^3 
+2(s-t)t W^2(W-1)^2}{(s-t)^2t^2 W^2(W-1)^2} W'  \nn \\ 
&&+\frac{n^2s^2W^2[(s-t)W-t]-m^2t^2(W-1)^2[(s-t)W+t-2s] 
- \mu^2(s-t)^2W^2(W-1)^2 [(s+t)W-t]}{2(s-t)^3t^3 W(W-1)}\ , \nn \\  
\label{eq:3dord}   
\eea
where the prime denotes differentiation with respect to $t$. 
Eq. (\ref{eq:3dord}) can be integrated once using the integrating 
factor 
\[ \frac{t(t-s)[(t-s)W-t]}{W(W-1)}   \] 
leading to the following second-order equation for $W$: 
\bea 
W''&=& \frac{1}{2}\left( \frac{1}{W} + \frac{1}{W-1} 
+\frac{t-s}{(t-s)W-t} \right) W^{\prime 2} 
- \left( \frac{1}{t} + \frac{1}{t-s} - \frac{W-1}{(t-s)W-t} \right) 
W'  \nn \\ 
&& + \frac{W(W-1)[(t-s)W-t]}{2t(t-s)}\left( \frac{\mu^2}{t} - 
\frac{m^2}{(t-s)W^2} + \frac{n^2 s}{t(t-s)(W-1)^2} 
- \frac{\nu^2 s}{((t-s)W-t)^2} \right) \nn \\    , 
\label{eq:2ndord} 
\eea 
where the integration constant is conveniently chosen as $\nu^2 s$ 
with $\nu$ independent of $s,t$  
(noting that $W$ depends on $s,t$ through the combination 
$s/t$). 
It is not hard to see that eq. (\ref{eq:2ndord}) is equivalent 
to the Painlev\'e VI equation (${\rm P}_{{\rm\small VI}}$)
after the trivial change of variables ~$t/(t-s)\mapsto \tau$~. 
Thus, we obtain  PVI in the standard form 
\begin{equation*}
\phantom{aaaaaaa}\begin{array}{ll}
\displaystyle\frac{d^2w}{d\tau^2}&\displaystyle=\frac{1}{2}\left(
\frac{1}{w}+\frac{1}{w-1}+
\frac{1}{w-\tau}\right)\left(\frac{dw}{d\tau}\right)^2 -
\left(\frac{1}{\tau}+\frac{1}{\tau-1}+\frac{1}{w-\tau}\right)
\frac{dw}{d\tau}\\ &\\
&\displaystyle\ + \frac{w(w-1)(w-\tau)}{2\tau^2(\tau-1)^2}\left( 
\mu^2-m^2\frac{\tau}{w^2} +n^2\frac{\tau-1}{(w-1)^2}-
\nu^2\frac{\tau(\tau-1)}{(w-\tau)^2} \right)\   . \\ 
\end{array}\ \quad {\rm P}_{{\rm\small VI}}
\end{equation*} 
for $w(\tau)=W(t)$. 
with the identification of the parameters as follows: 
\bea\label{eq:parms} 
\ar= \mu^2\     \ ,\      \ \bb= m^2\    \,\     \ \gm= n^2\      \ ,
\     \ \dd=\nu^2\      
\eea 
(we have rescaled the  
parameters compared with \cite{NJH}). 
 PVI, which incidentally was first found by R. Fuchs in \cite{Fuchs} in 
1905, (and not by either Painlev\'e or Gambier as is ocasionally 
claimed in the recent literature), was found to be connected to the 
lattice KdV systems in \cite{NRGO}. It is known that PVI is related 
to the Toda lattice, cf. e.g. \cite{Oka,Oka1}. In a recent paper 
\cite{NJH} we presented the full Miura 
chain associated with PVI, consisting of ordinary differential as 
well as ordinary difference equations, showing that in fact the 
parameter-family of ODEs that constitute PVI has a natural 
description  in terms of both discrete as well as continuous equations. 
The Miura chain for PVI is just the reduction of the Miura chain for the 
SPDE under the similarity constraint (\ref{eq:spdeconstr}), 
although we implimented this constraint differently in \cite{NJH}.  

Thus the SPDE (\ref{eq:SPDE}) yields PVI as similarity reduction 
with four arbitrary parameters. As far as we are aware 
this is the first case of an integrable scalar PDE that reduces 
to the full 
PVI equation. In the literature, cf. e.g. \cite{FLMS}-\cite{Tod} 
PVI was obtained in several circumstances, either from the reduction 
of the three-wave resonant interaction system or in the context of the 
Einstein equations. However, in these instances 
only special parameter-cases of PVI were obtained. In \cite{Halb} 
a connection between PVI and so-called generalised Ernst equations 
was studied, which seems to give a connection with full PVI, but 
the starting point is a complicated system of equations. What we have 
demonstrated here is that the full PVI equation 
is directly connected to the KdV 
hierarchy. 

\subsection*{Connection with Equations on the Lattice} 

We finish by pointing out that the SPDE (\ref{eq:SPDE}) as well 
as the other generating PDEs in the Miura chain are not only 
compatible as parameter-families of PDEs in the sense pointed 
out above, but also are compatible with a system of {\it discrete} 
equations. In fact, the actual derivation of the SPDE was based 
on this underlying structure which involves continuous as 
well discrete components. The following statement holds: 

\paragraph{Proposition:} The SPDE (\ref{eq:SPDE}) is consistent with 
the following set of differential-difference equations for 
$z(s,t)=z_{n,m}(s,t)$: 
\bse\label{eq:zdif}\bea 
-t\frac{\pl z_{n,m}}{\pl t}&=& n\frac{(z_{n+1,m}-z_{n,m})
(z_{n,m}-z_{n-1,m})}{z_{n+1,m}-z_{n-1,m}} \label{eq:zdifa}  \\ 
-s\frac{\pl z_{n,m}}{\pl s}&=& m\frac{(z_{n,m+1}-z_{n,m})
(z_{n,m}-z_{n,m-1})}{z_{n,m+1}-z_{n,m-1}} \label{eq:zdifb}\    ,   
\eea\ese 
in which the parameters of the SPDE play the role of independent 
discrete variables. 

Actually one can show by direct calculation that the operations of 
shifting in the variables $n$ and $m$ governed by the equations 
(\ref{eq:zdif}) commute with the continuous evolution according 
to the flows in terms of the variables $s$ and $t$. A similar statement 
holds for the other generating PDEs, where the differential-difference 
equations for $v(s,t)=v_{n,m}(s,t)$ read
\begin{equation}\label{eq:vdif}
-2t\frac{\pl}{\pl t} \log v_{n,m} = n a_{n,m}\     \ ,\      \  
-2s\frac{\pl}{\pl s} \log v_{n,m} = n b_{n,m}\   , 
\end{equation}
with the auxiliary variables $a(s,t)=a_{n,m}(s,t)$ and 
$b(s,t)=b_{n,m}(s,t)$ which appeared in the Lax representation 
(\ref{eq:Lax}) given explicitly in terms of $v_{n,m}$ 
by the relations 
\begin{equation} \label{eq:a-b}
a_{n,m}\equiv\frac{v_{n+1,m}-v_{n-1,m}}{v_{n+1,m}+v_{n-1,m}}
\     \ ,\     \
b_{n,m}\equiv\frac{v_{n,m+1}-v_{n,m-1}}{v_{n,m+1}+v_{n,m-1}}
 \   . \end{equation} 
Also for the object $U(s,t)=U_{n,m}(s,t)$ we have a set of 
differential-difference equations compatible with the 
generating PDE (\ref{eq:UPDE}), namely  
\begin{equation}\label{eq:udif}
\frac{\pl U_{n,m}}{\pl t}=\frac{n}{U_{n+1,m}-U_{n-1,m}}\     \ ,
\      \  
\frac{\pl U_{n,m}}{\pl s} = \frac{m}{U_{n,m+1}-U_{n,m-1}}\  . 
\end{equation}

Finally, to complete the picture we mention the associated fully 
discrete equation for the variable $z_{n,m}(s,t)$, i.e the partial 
{\it difference} equation (P$\Delta$E) 
\begin{equation}\label{eq:dSKdV}
\frac{(z_{n,m}-z_{n+1,m})(z_{n,m+1}-z_{n+1,m+1})}{(z_{n,m}-z_{n,m+1})
(z_{n+1,m}-z_{n+1,m+1})} = \frac{s}{t}\    .    
\end{equation}
The equation (\ref{eq:dSKdV}), which was first given in \cite{KDV}, is 
possibly the most fundamental equation in the entire structure, and 
it has been studied at length in the context of discrete conformal 
maps, see e.g. the recent monograph \cite{BS}. Note that in 
(\ref{eq:dSKdV}) the independent variables $s$,$t$ of the SPDE now 
enter as parameters of the discrete equation. In \cite{Dorf} (see  
also \cite{DIGP}) the similarity reduction on the lattice was 
formulated, namely by imposing the following constraint which is 
compatible on the lattice with the equation (\ref{eq:dSKdV}):  
\begin{equation} \label{eq:skdvconstr1}
n\frac{(z_{n+1,m}-z_{n,m})(z_{n,m}-z_{n-1,m})}{ z_{n+1,m}-z_{n-1,m}}
+ m \frac{ (z_{n,m+1}-z_{n,m})(z_{n,m}-z_{n,m-1})}{z_{n,m+1}-
z_{n,m-1} } = \bmu z_{n,m}\   ,
\end{equation}
It was, however only in the recent paper \cite{NJH} that 
a closed-form third-order ordinary difference equation (O$\Delta$E) 
was found that represents the similarity reduced equation, namely 
\bea\label{eq:dSPVI} 
&&(r^2-1)(z_{n+1}-z_n)^2  = \nn\\ 
&&= \left[ 2r^2\frac{\mu z_{n+1}(z_{n+2}-z_n)-(n+1)
(z_{n+2}-z_{n+1})(z_{n+1}-z_n)}{(m-\mu-\nu)(z_{n+2}-z_n)+(n+1)
(z_{n+2}-2z_{n+1}+z_n)}+z_{n+1}-z_n\right] \times \nn\\ 
&&~~~~\times\left[ 2r^2\frac{\mu z_n(z_{n+1}-z_{n-1})-n
(z_{n+1}-z_n)(z_n-z_{n-1})}{(m-\mu+\nu)(z_{n+1}-z_{n-1})+n
(z_{n+1}-2z_n+z_{n-1})}+z_n-z_{n+1}\right] \   .  \nn\\ 
\eea
Eq. (\ref{eq:dSPVI}) is related to the {\it discrete Painlev\'e equation} 
that was studied at length in \cite{NRGO}. Discrete Painlev\'e form 
an exciting class of ordinary difference equations which is becoming 
increasingly an area of intense activity (see \cite{Carg} for a 
recent review). 
In \cite{NJH} we presented the discrete and continuous Miura chains 
of both discrete and continuous equations of Painlev\'e type that are 
associated with the third-order O$\Delta$E (\ref{eq:dSPVI}). As is 
obvious from what we said above on the similarity reduction of the 
SPDE to the full PVI equation, these discrete and continuous 
equations are directly associated with PVI.

\section{Conclusions and Outlook} 

It has been a long-standing conjecture whether all differential 
equations of Painlev\'e type could be derived as reductions from 
a larger integrable nonlinear equation or system of PDE's by 
similarity reduction. This is, in a sense, the converse of the 
famous ARS hypothesis \cite{ARS}. The programme developed in 
the papers \cite{Dorf,DIGP,NRGO,NJH} has demonstrated that 
there is an intimate interplay between discrete and continuous structures 
behind the Painlev\'e equations and that notably PVI and its 
discrete counterparts are obtainable from either a system of 
P$\Delta$Es on the full lattice or of a very rich parameter-family 
of PDEs in the continuum, which we  have exhibited in the present 
paper. An obvious conjecture is that these connections also extend to 
discrete Painlev\'e equations which are of a possibly richer type, 
in particular the so-called $q$-Painlev\'e equations, which seemingly 
are ``beyond'' the PVI equation. In that case, we may expect that there 
exist partial difference equations of $q$-type, generalizing the 
SPDE (\ref{eq:SPDE}), whose similarity reduction leads to those 
discrete Painlev\'e equations. The ultimate aim would be to find 
the ``big equation'' that sits above the recently found most 
general discrete Painlev\'e equation by Sakai, \cite{Sakai}, which would be 
a partial difference equation of elliptic type.

\end{document}